\documentclass[10pt,aps,twocolumn,prl,showpacs]{revtex4-1}
\usepackage[normalem]{ulem}
\usepackage[usenames]{color}
\usepackage{graphicx}
\usepackage{ulem}
\usepackage{amsfonts}
\usepackage{amsmath}
\begin{document}
\title{High-Precision Spectroscopy with Counter-Propagating Femtosecond Pulses}
\author{Itan Barmes}
\author{Stefan Witte}
\author{Kjeld S. E. Eikema}\email{k.s.e.eikema@vu.nl}
\affiliation{Department of Physics and Astronomy, LaserLaB, VU University, de Boelelaan 1081, 1081 HV Amsterdam, The Netherlands}

\begin{abstract}
An experimental realization of high-precision direct frequency comb spectroscopy using counter-propagating femtosecond pulses on two-photon atomic transitions is presented. Doppler broadened background signal, hampering precision spectroscopy with ultrashort pulses, is effectively eliminated with a simple pulse shaping method. As a result, all four 5S-7S two-photon transitions in a rubidium vapor are determined with both statistical and systematic uncertainties below 10$^{-11}$, which is an order of magnitude better than previous experiments on these transitions.
\end{abstract}
\maketitle

One of the hallmarks of laser spectroscopy has been the theoretical prediction~\cite{Chebotayev1970} and the experimental realization~\cite{Bloembergen1974a,HANSCH1974} of two-photon Doppler-free spectroscopy using continuous wave (CW) lasers in a counter-propagating beam geometry. In this method, the Doppler shift due to the velocity of an atom in the lab frame is compensated by an opposite shift from a counter-propagating beam. Therefore Doppler-free signals can be obtained, even without the use of laser cooling and trapping techniques. A classic example of Doppler-free two-photon excitation with high accuracy is 1S-2S spectroscopy in hydrogen~\cite{Parthey2011}. Doppler-free two-photon spectroscopy has been essential in the determination of the Rydberg constant and proton charge radius~\cite{Mohr2008}, accurate tests of quantum electrodynamics, and the detection of possible drifts in fundamental constants~\cite{Fischer2004}. An extension of the Doppler-free method to nanosecond pulses~\cite{HaenschPRL1975} has been implemented for high-precision spectroscopy in, for example, molecular hydrogen~\cite{Dickenson2013} and muonium~\cite{Meyer2000}. A more recent development in precision spectroscopy is the realization of the optical frequency comb, which revolutionized the field of precision measurements~\cite{Rosenband2008,Coddington2009}. An optical frequency comb is based on the precise phase relation of a train of ultrashort pulses and acts as a frequency ruler, connecting the rf and optical frequency domains. In the field of precision spectroscopy, optical frequency combs were initially used as a referencing tool for a separate CW excitation laser. Subsequently, frequency combs were used to induce transitions directly for precision measurements~\cite{Marian2004,Gerginov2005,Wolf2009}, which marked the beginning of a new field of direct frequency comb spectroscopy (DFCS). The high peak intensity of ultrashort pulses from frequency comb lasers also facilitates frequency conversion via nonlinear processes, paving the way for high-precision spectroscopy in wavelength regions where CW lasers do not exist~\cite{Kandula2010,Witte2005,Cingoz2012,Diddams2007}.

Combining DFCS with Doppler reduction using counter propagating beams has therefore drawn significant attention. One approach is to drive the transition via an intermediate resonance (stepwise excitation) which enhances signal strength~\cite{Bjorkholm1976,Stalnaker2010}. The signal is then indeed free of Doppler broadening, however, due to the imbalance between the two frequencies the line center is shifted. Furthermore, population transfer to the intermediate level complicates assessment of systematic effects. Alternatively, non-resonant excitation is also possible on two-photon transitions. Pairs of modes from the comb laser can then combine to the same total energy, so that the full comb spectrum contributes to the signal~\cite{Baklanov1977,Fendel2007}. In this scheme, Doppler-free excitation only occurs in the region of space where the counter-propagating pulses overlap. For pulses of typical frequency comb lasers with a duration in the femtosecond range, this zone is limited to tens of micrometers in length, while Doppler-broadened excitation with co-propagating photons can take place over the whole beam path. Therefore, a dominating and detrimental background of Doppler-broadened signal impairs high-precision DFCS. Recently it was shown that the Doppler-broadened background can be reduced by stretching the pulses with group-velocity dispersion~\cite{Ozawa2012}, and that it can even be completely eliminated using concepts from quantum coherent control~\cite{Barmes2012}. In the latter case sophisticated pulse shaping techniques with a spatial light modulator were employed.

In this letter, we demonstrate a general method that enables high-resolution DFCS on two-photon transitions in a counter-propagating geometry. We introduce a simple and flexible split-pulse technique to eliminate the Doppler-broadened background. Combined with the versatility of DFCS we acquire signal with excellent signal-to-noise ratio (SNR) and low sensitivity to systematic effects. We demonstrate the possibilities of this method by performing absolute frequency measurements on the 5S-7S transitions in rubidium. The resulting accuracy of the 4 measured transitions (2 hyperfine transitions in 2 Rb isotopes) is an order of magnitude better than previous demonstrations with either DFCS~\cite{Marian2005} or CW lasers~\cite{Chui2005}.

\begin{figure}
\includegraphics[width=\columnwidth]{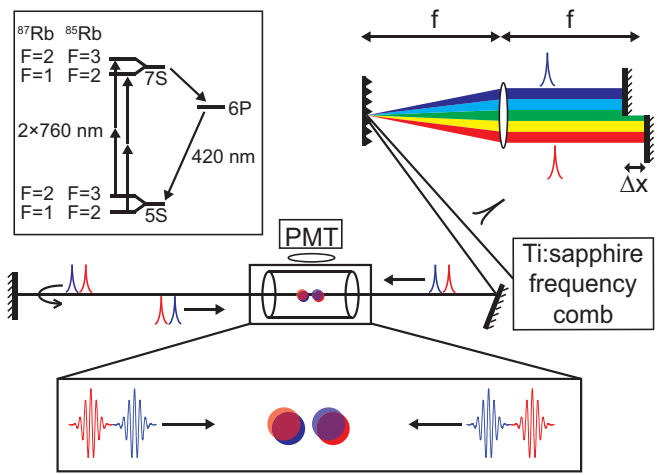}
\caption{A schematic of the shaping and spectroscopy setup. Each frequency comb pulse is split into a red and a blue sub-pulse in a simplified shaping apparatus. A delay between the sub-pulses reduces single-sided excitation while not affecting the total counter-propagating signal (localized in two separate excitation regions). The inset shows a simplified level scheme of atomic rubidium.}
\label{fig:setup}
\end{figure}

The experimental setup is shown schematically in~Fig.~\ref{fig:setup}. The frequency comb used in this experiment is based on a mode-locked Ti:sapphire oscillator with a pulse repetition rate that can be tuned between 140 and 180~MHz. It has a central wavelength of 760 nm, and a full width half maximum (FWHM) bandwidth of approximately 40 nm. The spectrum is composed of a large collection of equidistant narrow modes which are described by the comb equation $f_n=f_0+n\times f_{rep}$. Here $f_{0}$ is the carrier-envelope offset frequency, $f_{rep}$ is the repetition frequency, and $n$ is an integer mode number with a typical value of 10$^6$. Both comb parameters ($f_{rep}$ and $f_0$) are locked to low-noise rf generators, which themselves are referenced to a GPS-disciplined Rb atomic clock (better than $2\times 10^{-12}$ fractional accuracy).

Both the ground (5S) and excited (7S) states are split due to the hyperfine interaction. Selection rules dictate that only transitions between levels with the same hyperfine quantum number ($\Delta$F=0) are allowed. A simplified level structure of rubidium is shown as an inset of Fig.~\ref{fig:setup}. The spectroscopy is conducted in a commercial glass cell containing the two stable isotopes $^{85}$Rb and $^{87}$Rb. The transitions are induced by focusing frequency comb pulses in the middle of the cell with f=150 mm lenses to a beam size of about 100~$\mu$m at the focus. A mirror reflects the pulses back so that consecutive pulses overlap at the focus. Excitation to the 7S is monitored by detecting the 420~nm fluorescence from cascade decay via the 6P state with a photomultiplier tube (PMT).

Atomic excitation with counter-propagating femtosecond pulses presents a challenge as the (Doppler broadened) single-sided signal is not confined to the small overlap region and will therefore obscure the counter-propagating signal. To eliminate this background signal we apply a group delay between the lower and upper half of the spectrum, effectively splitting each pulse into a "red" and a "blue" sub-pulse with a relative delay on a picosecond time scale. As a combination of red and blue sub-pulses are simultaneously needed to induce the transition, increasing the time delay between them will diminish the single-sided signal. We achieve this situation with a particularly simple pulse shaper configuration, consisting of a grating, a lens and two mirrors (see~Fig.~\ref{fig:setup}). Each frequency comb pulse is spatially dispersed using the first half of a zero-dispersion 2f-2f configuration. The laser light is reflected back at the Fourier plane using two separate mirrors to form the desired red and blue sub-pulses, each containing half of the original spectrum. The time delay between the two sub-pulses is adjusted by displacing one of the two mirrors, while the bandwidth of each sub-pulse can be controlled by placing a hard aperture in the Fourier plane (not shown in the figure). Throughout the measurements the laser spectrum did not exceed 40~nm in order to avoid the single-photon excitation to the 5p state at 780~nm.

\begin{figure}
\includegraphics[width=\columnwidth]{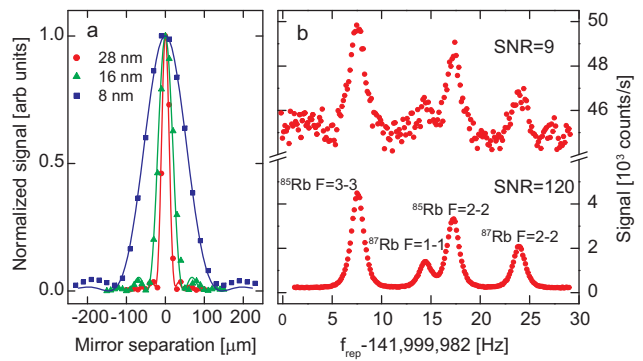}
\caption{(a) Normalized excitation rates of single-sided signal as a function of the mirror separation in the shaping setup. A suppression of better than 98\% is achieved for the various spectral widths. (b) A scan over all 4 5S-7S transitions, with (lower curve) and without (upper curve) background suppression, demonstrating the dramatic improvement in SNR. The SNR of the background-free measurement is only limited by shot-noise fluctuations. The traces were taken with a spectral bandwidth of 28~nm and an integration time of 3 seconds per point.}
\label{fig:background_reduction}
\end{figure}

We test this principle by blocking the returning beam in Fig.~\ref{fig:setup}, so that only single-sided excitation is induced. The Doppler width of the relevant transitions at room temperature is about 1~GHz, which is much larger than the spacing between the comb modes. This washes out the comb structure and the resulting signal generates a background that is independent of the comb parameters $f_0$ and $f_{rep}$. In Fig.~\ref{fig:background_reduction}a the single-sided signal is plotted as a function of the time delay between the sub-pulses for various spectral widths, together with numerical simulations (the solid lines) based on a framework developed in~\cite{Meshulach1999}. A larger temporal separation is needed for elimination of the single-sided signal when the pulses have a smaller bandwidth (because the pulse duration is then longer). This measurement can be seen as a type of cross correlation between the two sub-pulses. However, it is important to note that the two-photon signal does not simply depend on the pulse duration and intensity. For example, adding higher odd-order dispersion lengthens the pulses but does not change the total two-photon signal~\cite{Meshulach1998,Barmes2012}. For the investigated spectral bandwidths we find a background reduction of at least 98\% (this number is limited by measurement noise), for a mirror separation of less than 150~$\mu m$. A small mirror separation is advantageous in order to prevent deformation of the laser beam after the shaping apparatus.

With the single-sided signal effectively eliminated, we add the back-reflected beam in~Fig.~\ref{fig:setup}. Red and blue sub-pulses from opposite directions now overlap in two separate spatial regions and induce counter-propagating signal. The counter-propagating beam geometry reduces the original 1~GHz Doppler width to below the value of $f_{rep}$ so that excitation only takes place if combinations of modes are resonant with the transition frequency $f_t$. A scan over the various 5S-7S transitions is achieved by taking small steps of the repetition frequency $f_{rep}$. Such a scan is presented in Fig.~\ref{fig:background_reduction}b where a significant improvement in SNR is clearly visible when single-sided excitation is eliminated.

\begin{figure}
\includegraphics[width=\columnwidth]{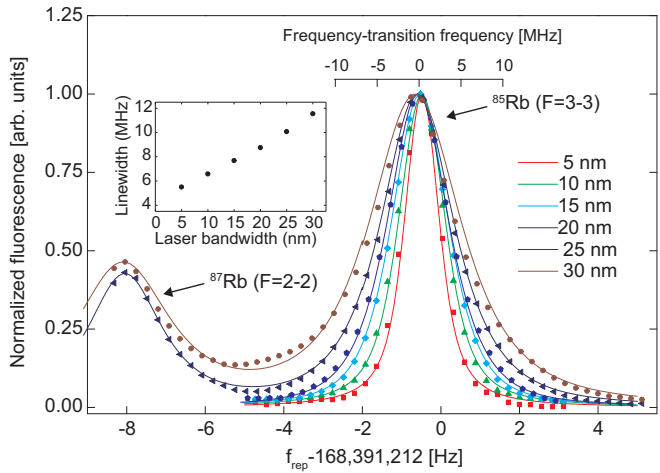}
\caption{Multiple scans of background-free signal over a single transition. The absolute frequency scale presented above the traces is calculated from the comb equation. The solid lines are computed from Eq.~\ref{eq:Doppler_profile} to show the validity of the line shape model. In these calculations a single Gaussian wavelength-dependent scaling function (FWHM 20~nm) of the intensity was used to account for chromatic aberration due to the lenses used in the experimental setup. Excitation with a larger bandwidth leads to a broader line width, which results in less accurate results and a possible systematic shift due to the overlap with neighboring transitions. A larger bandwidth also has more optical power, which shifts the transition due to the AC Stark effect. The linear dependence between laser bandwidth and residual Doppler broadening is shown in the inset.}
\label{fig:linewidth_vs_spectrum}
\end{figure}

\begin{figure}
\includegraphics[width=\columnwidth]{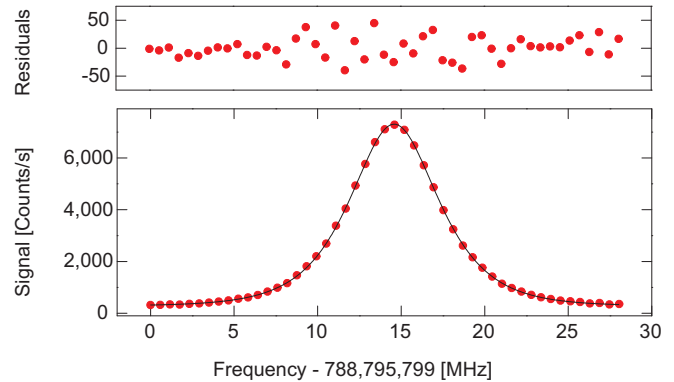}
\caption{A typical recording of the $^{85}$Rb(F=3-3) transition. The repetition rate of the FC is chosen such that this line has a maximum distance to the other 3 transitions. The excellent SNR allows determination of the line center to about 1:1000 of the measured line width}
\label{fig:fit_with_residuals}
\end{figure}

The peaks in Fig.~\ref{fig:background_reduction}b correspond to values of the comb parameters ($f_{rep}$, $f_0$ and n) for which $f_t/2$ coincides with one of the comb modes or is exactly between two modes. As a consequence, a scan of f$_{rep}$ results in a periodic signal with periodicity of f$_{rep}/2$ (for an overview of DFCS see~\cite{Stowe2008a}). Furthermore, whenever two photons of a single mode sum up to the transition frequency ($f_t=f_n+f_n$) then other pairs of modes are also resonant ($f_{t}=f_{n-k}+f_{n+k}$), which means that all of the comb modes participate in the excitation. For each pair of frequencies (f$_1$,f$_2$) the line shape can be described as a Voigt profile (convolution of a Gaussian g$_D$ and a Lorentzian g$_b$), with a Gaussian width of $2\sqrt{\ln 2}\frac{u}{c}|$f$_1$-f$_2|$. $u=\sqrt{2 k_B T/M}$ is the most probable velocity of atoms with mass M at temperature T. The line profile in this situation is equal to~\cite{Bjorkholm1976}:

\begin{equation}
\begin{split}
|a_f^{(2)}|^2\propto& \left ( \frac{|E(f_1)|^2|E(f_2)|^2}{(f_1-f_i)^2((f_t-f_1-f_2)^2+1/4 \tau_f^2)} \right ) \ast g_D,\\
g_D=&\exp\left[-\left(\frac{c}{u}\right)^2\left(\frac{f_t-(f_1+f_2)}{f_1-f_2}\right)^2\right],
\end{split}
\label{eq:Doppler_profile_two_frequencies}
\end{equation}
where \textit{E} is the spectral amplitue and $\tau_f$ is the decay time from the excited state. In this derivation it was assumed that no intermediate levels (f$_i$) are populated. Extending this equation to account for all possible mode combinations is achieved by replacing the single frequencies (f$_1$,f$_2$) with the comb equation and summing over all comb modes. This leads to the following equation:

\begin{equation}
\begin{split}
|a_f^{(2)}|^2&(f_{rep})\propto\\
\sum_{n_1,n_2}& \left ( \frac{|E(f_{n_1})|^2 |E(f_{n_2})|^2}{(f_{n_1}-f_i)^2((f_t-f_{n_1}-f_{n_2})^2+1/4 \tau_f^2)} \right )\ast g_D,\\
g_D=&\exp\left[-\left(\frac{c}{u}\right)^2\left(\frac{f_t-(f_{n_1}+f_{n_2})}{f_{n_1}-f_{n_2}}\right)^2\right].
\end{split}
\label{eq:Doppler_profile}
\end{equation}
The experimental results of the line profile of the $^{85}$Rb (F=3-3) transition are shown in Fig~\ref{fig:linewidth_vs_spectrum}. As predicted by Eq.~\ref{eq:Doppler_profile}, the line width is proportional to the laser bandwidth. For bandwidths larger than 25 nm a neighboring transition $^{87}$Rb (F=2-2) starts to overlap with the $^{85}$Rb (F=3-3) line shape. The solid lines in Fig.~3 are numerical calculations of Eq.~\ref{eq:Doppler_profile} for different spectral bandwidths, while transit-time broadening is incorporated in $\tau_f$. The exact line shape is sensitive to additional experimental conditions. For example, chromatic aberrations due to the various lenses in the setup need to be accounted for as the intensity at the focus is wavelength dependent (see caption Fig.~\ref{fig:linewidth_vs_spectrum}). Using Eq.~\ref{eq:Doppler_profile} as a fitting function is cumbersome for determining the line center. However, Eq.~\ref{eq:Doppler_profile} is a symmetric function. Therefore, fitting any other symmetric function to the data does not introduce a systematic error in determining the line center. For this purpose we have used a simplified fitting function consisting of a sum of a single Gaussian and Lorentzian for each transition. In this model the widths of the Gaussian and Lorentzian functions are given as free parameters and are not physically meaningful. Nevertheless, this approach is computationally very fast and we verified that it does not lead to a systematic shift in the determination of the line center. A typical data set including the fitting function and fit residuals is shown in Fig.~\ref{fig:fit_with_residuals}. This trace was recorded with a laser bandwidth of 10~nm, which gives the best compromise between signal strength and residual Doppler broadening. The measured transition line width was 6~MHz FWHM (comparable to the 1.8~MHz natural line width), and the SNR allows a determination of the line center to better than 1:1000 of the measured line width, limited by shot noise fluctuations. The structureless residuals validate that our model function is successful in accurately determining the line center.

Before an absolute frequency determination of the individual transitions can be made, all possible systematic shifts need to be quantified and corrected for. Due to the low pulse energy (30-300 pJ) and peak intensity ($<$50 MW/cm$^2$) strong field effects such as multiphoton ionization and self-phase modulation are negligible. The main systematic effects in the present work are pressure effects, magnetic (Zeeman) shift, and AC Stark shifts. Pressure shifts can manifest in two different ways. First, collisions between Rb atoms can shift the transition frequency as a linear function of the pressure in the vapor cell. Previous studies of this effect have shown that the pressure shift is equal to -103.4(10.0)~kHz/mTorr~\cite{Chui2005}. As the pressure in our experiment was kept below 2$\times$10$^{-5}$ mTorr, a shift of less than 2~kHz is expected. Impurities in the vapor cell can also lead to systematic shifts. This is more difficult to quantify as the pressure of impurities is hardly affected by changes in the temperature. We take a conservative upper limit for the pressure shift equal to 5~kHz. The shift due to external magnetic fields is small for the measured S to S transitions, as the linear Zeeman shift is zero. However, second-order Zeeman shift of a few kHz/G$^2$ needs to be taken into account. We apply a correction for the measured transition frequencies derived from calculations of the second-order Zeeman shift~\cite{Li2009} due to the uncompensated earth's magnetic field. This shift is different for each transition, ranging from 0.5 to 1.2~kHz.

An additional systematic effect is due to the presence of a light field (AC Stark shift). This shift scales linearly with the average power of the laser~\cite{Fendel2007}. In order to correct for this shift (a few kHz/mW for our experimental conditions) we have performed measurements at different optical powers and extrapolated to zero. This was done for 10 measurement sets of the 5S-7S (F=3-3) in $^{85}$Rb, leading to an absolute transition frequency of~788~795~814~061.8~kHz with statistical and systematic uncertainties of 4 and 5 kHz respectively. The transition frequency was corrected for the above mentioned systematic shifts, including smaller corrections for the second-order Doppler shift (-420~Hz at 60$^\circ$C) and black body radiation shift (-630~Hz at 60$^\circ$C). The statistical accuracy is an order of magnitude better than in previous studies of this transition~\cite{Chui2005}.

We have also performed measurements of difference frequencies between the various hyperfine transitions by scanning over all four transitions (Fig.~\ref{fig:background_reduction}b) and extracting the difference frequencies. As both AC Stark shift and pressure shift are the same for all 4 transitions, the difference frequencies are insensitive to the laser intensity and gas pressure. The leading systematic uncertainty is then the second-order Zeeman shift which is corrected for in the same way as described above. By combining these relative measurements with the accurately calibrated $^{85}$Rb (F=3-3) transition, we have determined the absolute frequencies of all 4 transition, as well as the hyperfine A coefficients and the isotope shift of the upper states (the values of the ground state splittings are taken from~\cite{Arimondo1977}). The final results are summarized in Table~\ref{tab:table1}.

\begin{table}
\caption{\label{tab:table1} A summary of the spectroscopy results. The values in parentheses are the statistical and systematic uncertainties. All values are given in kHz.}
\begin{ruledtabular}
\begin{tabular}{lr}
Transitions\\
\hline
$^{85}$Rb$(F=3-3)$ & $788~795~814~061.8(4.0)_{stat}(5.1)_{sys}$\\
$^{85}$Rb$(F=2-2)$ & $788~798~565~752.1(6.4)_{stat}(5.6)_{sys}$\\
$^{87}$Rb$(F=2-2)$ & $788~794~768~940.1(7.2)_{stat}(5.2)_{sys}$\\
$^{87}$Rb$(F=1-1)$ & $788~800~964~119.7(7.2)_{stat}(5.4)_{sys}$\\
\hline
Hyperfine A constants\\
\hline
$^{85}$Rb 7S & $ 94~680.7(3.0)_{stat}(2.1)_{sys}$\\
$^{87}$Rb 7S & $ 319~751.8(5.0)_{stat}(0.9)_{sys}$\\
\hline
Upper state isotope shift\\
\hline
$^{85}$Rb -$^{87}$Rb & $ 131~533.2(12.1)_{stat}(8.5)_{sys}$\\
\end{tabular}
\end{ruledtabular}
\end{table}

In conclusion, we have demonstrated the elimination of Doppler-broadened background, using a simple shaping setup consisting of a grating and two mirrors, which enables high-precision spectroscopy with fs pulses in a counter-propagating beam geometry. DFCS on room-temperature Rb gas was performed with up to a tenfold improvement of the absolute frequencies of 4 two-photon transitions. This technique provides a simple and robust method for high-precision spectroscopy using a single laser. The method is also compatible with XUV comb generation, which opens the perspective of Doppler-reduced two-photon precision measurements in the XUV.

\begin{acknowledgments}
{S.W. acknowledges support from the Netherlands Organization for Scientific Research (NWO Veni grant 680-47-402). K.S.E.E. acknowledges support from the NWO (Vici grant no. 680-47-310), the Foundation for Fundamental Research on Matter (FOM) through its programme 'Broken Mirrors and Drifting Constants', and Laserlab Europe (JRA INREX).}
\end{acknowledgments}
\bibliography{biblio}

\begin{thebibliography}{31}%
\makeatletter
\providecommand \@ifxundefined [1]{%
 \@ifx{#1\undefined}
}%
\providecommand \@ifnum [1]{%
 \ifnum #1\expandafter \@firstoftwo
 \else \expandafter \@secondoftwo
 \fi
}%
\providecommand \@ifx [1]{%
 \ifx #1\expandafter \@firstoftwo
 \else \expandafter \@secondoftwo
 \fi
}%
\providecommand \natexlab [1]{#1}%
\providecommand \enquote  [1]{``#1''}%
\providecommand \bibnamefont  [1]{#1}%
\providecommand \bibfnamefont [1]{#1}%
\providecommand \citenamefont [1]{#1}%
\providecommand \href@noop [0]{\@secondoftwo}%
\providecommand \href [0]{\begingroup \@sanitize@url \@href}%
\providecommand \@href[1]{\@@startlink{#1}\@@href}%
\providecommand \@@href[1]{\endgroup#1\@@endlink}%
\providecommand \@sanitize@url [0]{\catcode `\\12\catcode `\$12\catcode
  `\&12\catcode `\#12\catcode `\^12\catcode `\_12\catcode `\%12\relax}%
\providecommand \@@startlink[1]{}%
\providecommand \@@endlink[0]{}%
\providecommand \url  [0]{\begingroup\@sanitize@url \@url }%
\providecommand \@url [1]{\endgroup\@href {#1}{\urlprefix }}%
\providecommand \urlprefix  [0]{URL }%
\providecommand \Eprint [0]{\href }%
\providecommand \doibase [0]{http://dx.doi.org/}%
\providecommand \selectlanguage [0]{\@gobble}%
\providecommand \bibinfo  [0]{\@secondoftwo}%
\providecommand \bibfield  [0]{\@secondoftwo}%
\providecommand \translation [1]{[#1]}%
\providecommand \BibitemOpen [0]{}%
\providecommand \bibitemStop [0]{}%
\providecommand \bibitemNoStop [0]{.\EOS\space}%
\providecommand \EOS [0]{\spacefactor3000\relax}%
\providecommand \BibitemShut  [1]{\csname bibitem#1\endcsname}%
\let\auto@bib@innerbib\@empty
\bibitem [{\citenamefont {Vasilenko}\ \emph {et~al.}(1970)\citenamefont
  {Vasilenko}, \citenamefont {Chebotayev},\ and\ \citenamefont
  {Shishaev}}]{Chebotayev1970}%
  \BibitemOpen
  \bibfield  {author} {\bibinfo {author} {\bibfnamefont {L.}~\bibnamefont
  {Vasilenko}}, \bibinfo {author} {\bibfnamefont {V.}~\bibnamefont
  {Chebotayev}}, \ and\ \bibinfo {author} {\bibfnamefont {A.}~\bibnamefont
  {Shishaev}},\ }\href@noop {} {\bibfield  {journal} {\bibinfo  {journal} {Jetp
  Letters}\ }\textbf {\bibinfo {volume} {12}},\ \bibinfo {pages} {113}
  (\bibinfo {year} {1970})}\BibitemShut {NoStop}%
\bibitem [{\citenamefont {Levenson}\ and\ \citenamefont
  {Bloembergen}(1974)}]{Bloembergen1974a}%
  \BibitemOpen
  \bibfield  {author} {\bibinfo {author} {\bibfnamefont {M.}~\bibnamefont
  {Levenson}}\ and\ \bibinfo {author} {\bibfnamefont {N.}~\bibnamefont
  {Bloembergen}},\ }\href {\doibase 10.1103/PhysRevLett.32.645} {\bibfield
  {journal} {\bibinfo  {journal} {Physical Review Letters}\ }\textbf {\bibinfo
  {volume} {32}},\ \bibinfo {pages} {645} (\bibinfo {year} {1974})}\BibitemShut
  {NoStop}%
\bibitem [{\citenamefont {H\"{a}nsch}\ \emph {et~al.}(1974)\citenamefont
  {H\"{a}nsch}, \citenamefont {Harvey}, \citenamefont {Meisel},\ and\
  \citenamefont {Schawlow}}]{HANSCH1974}%
  \BibitemOpen
  \bibfield  {author} {\bibinfo {author} {\bibfnamefont {T.}~\bibnamefont
  {H\"{a}nsch}}, \bibinfo {author} {\bibfnamefont {K.}~\bibnamefont {Harvey}},
  \bibinfo {author} {\bibfnamefont {G.}~\bibnamefont {Meisel}}, \ and\ \bibinfo
  {author} {\bibfnamefont {A.}~\bibnamefont {Schawlow}},\ }\href {\doibase
  10.1016/0030-4018(74)90331-9} {\bibfield  {journal} {\bibinfo  {journal}
  {Optics Communications}\ }\textbf {\bibinfo {volume} {11}},\ \bibinfo {pages}
  {50} (\bibinfo {year} {1974})}\BibitemShut {NoStop}%
\bibitem [{\citenamefont {Parthey}\ \emph {et~al.}(2011)\citenamefont
  {Parthey}, \citenamefont {Matveev}, \citenamefont {Alnis}, \citenamefont
  {Bernhardt}, \citenamefont {Beyer}, \citenamefont {Holzwarth}, \citenamefont
  {Maistrou}, \citenamefont {Pohl}, \citenamefont {Predehl}, \citenamefont
  {Udem}, \citenamefont {Wilken}, \citenamefont {Kolachevsky}, \citenamefont
  {Abgrall}, \citenamefont {Rovera}, \citenamefont {Salomon}, \citenamefont
  {Laurent},\ and\ \citenamefont {H\"{a}nsch}}]{Parthey2011}%
  \BibitemOpen
  \bibfield  {author} {\bibinfo {author} {\bibfnamefont {C.~G.}\ \bibnamefont
  {Parthey}}, \bibinfo {author} {\bibfnamefont {A.}~\bibnamefont {Matveev}},
  \bibinfo {author} {\bibfnamefont {J.}~\bibnamefont {Alnis}}, \bibinfo
  {author} {\bibfnamefont {B.}~\bibnamefont {Bernhardt}}, \bibinfo {author}
  {\bibfnamefont {A.}~\bibnamefont {Beyer}}, \bibinfo {author} {\bibfnamefont
  {R.}~\bibnamefont {Holzwarth}}, \bibinfo {author} {\bibfnamefont
  {A.}~\bibnamefont {Maistrou}}, \bibinfo {author} {\bibfnamefont
  {R.}~\bibnamefont {Pohl}}, \bibinfo {author} {\bibfnamefont {K.}~\bibnamefont
  {Predehl}}, \bibinfo {author} {\bibfnamefont {T.}~\bibnamefont {Udem}},
  \bibinfo {author} {\bibfnamefont {T.}~\bibnamefont {Wilken}}, \bibinfo
  {author} {\bibfnamefont {N.}~\bibnamefont {Kolachevsky}}, \bibinfo {author}
  {\bibfnamefont {M.}~\bibnamefont {Abgrall}}, \bibinfo {author} {\bibfnamefont
  {D.}~\bibnamefont {Rovera}}, \bibinfo {author} {\bibfnamefont
  {C.}~\bibnamefont {Salomon}}, \bibinfo {author} {\bibfnamefont
  {P.}~\bibnamefont {Laurent}}, \ and\ \bibinfo {author} {\bibfnamefont
  {T.~W.}\ \bibnamefont {H\"{a}nsch}},\ }\href {\doibase
  10.1103/PhysRevLett.107.203001} {\bibfield  {journal} {\bibinfo  {journal}
  {Physical Review Letters}\ }\textbf {\bibinfo {volume} {107}},\ \bibinfo
  {pages} {203001} (\bibinfo {year} {2011})}\BibitemShut {NoStop}%
\bibitem [{\citenamefont {Mohr}\ \emph {et~al.}(2008)\citenamefont {Mohr},
  \citenamefont {Taylor},\ and\ \citenamefont {Newell}}]{Mohr2008}%
  \BibitemOpen
  \bibfield  {author} {\bibinfo {author} {\bibfnamefont {P.~J.}\ \bibnamefont
  {Mohr}}, \bibinfo {author} {\bibfnamefont {B.~N.}\ \bibnamefont {Taylor}}, \
  and\ \bibinfo {author} {\bibfnamefont {D.~B.}\ \bibnamefont {Newell}},\
  }\href {\doibase 10.1103/RevModPhys.80.633} {\bibfield  {journal} {\bibinfo
  {journal} {Reviews of Modern Physics}\ }\textbf {\bibinfo {volume} {80}},\
  \bibinfo {pages} {633} (\bibinfo {year} {2008})}\BibitemShut {NoStop}%
\bibitem [{\citenamefont {Fischer}\ \emph {et~al.}(2004)\citenamefont
  {Fischer}, \citenamefont {Kolachevsky}, \citenamefont {Zimmermann},
  \citenamefont {Holzwarth}, \citenamefont {Udem}, \citenamefont {H\"{a}nsch},
  \citenamefont {Abgrall}, \citenamefont {Grunert}, \citenamefont {Maksimovic},
  \citenamefont {Bize}, \citenamefont {Marion}, \citenamefont {Santos},
  \citenamefont {Lemonde}, \citenamefont {Santarelli}, \citenamefont {Laurent},
  \citenamefont {Clairon}, \citenamefont {Salomon}, \citenamefont {Haas},
  \citenamefont {Jentschura},\ and\ \citenamefont {Keitel}}]{Fischer2004}%
  \BibitemOpen
  \bibfield  {author} {\bibinfo {author} {\bibfnamefont {M.}~\bibnamefont
  {Fischer}}, \bibinfo {author} {\bibfnamefont {N.}~\bibnamefont
  {Kolachevsky}}, \bibinfo {author} {\bibfnamefont {M.}~\bibnamefont
  {Zimmermann}}, \bibinfo {author} {\bibfnamefont {R.}~\bibnamefont
  {Holzwarth}}, \bibinfo {author} {\bibfnamefont {T.}~\bibnamefont {Udem}},
  \bibinfo {author} {\bibfnamefont {T.~W.}\ \bibnamefont {H\"{a}nsch}},
  \bibinfo {author} {\bibfnamefont {M.}~\bibnamefont {Abgrall}}, \bibinfo
  {author} {\bibfnamefont {J.}~\bibnamefont {Grunert}}, \bibinfo {author}
  {\bibfnamefont {I.}~\bibnamefont {Maksimovic}}, \bibinfo {author}
  {\bibfnamefont {S.}~\bibnamefont {Bize}}, \bibinfo {author} {\bibfnamefont
  {H.}~\bibnamefont {Marion}}, \bibinfo {author} {\bibfnamefont {F.~P.~D.}\
  \bibnamefont {Santos}}, \bibinfo {author} {\bibfnamefont {P.}~\bibnamefont
  {Lemonde}}, \bibinfo {author} {\bibfnamefont {G.}~\bibnamefont {Santarelli}},
  \bibinfo {author} {\bibfnamefont {P.}~\bibnamefont {Laurent}}, \bibinfo
  {author} {\bibfnamefont {A.}~\bibnamefont {Clairon}}, \bibinfo {author}
  {\bibfnamefont {C.}~\bibnamefont {Salomon}}, \bibinfo {author} {\bibfnamefont
  {M.}~\bibnamefont {Haas}}, \bibinfo {author} {\bibfnamefont {U.~D.}\
  \bibnamefont {Jentschura}}, \ and\ \bibinfo {author} {\bibfnamefont {C.~H.}\
  \bibnamefont {Keitel}},\ }\href {\doibase 10.1103/PhysRevLett.92.230802}
  {\bibfield  {journal} {\bibinfo  {journal} {Physical Review Letters}\
  }\textbf {\bibinfo {volume} {92}},\ \bibinfo {pages} {230802} (\bibinfo
  {year} {2004})}\BibitemShut {NoStop}%
\bibitem [{\citenamefont {H\"{a}nsch}\ \emph {et~al.}(1975)\citenamefont
  {H\"{a}nsch}, \citenamefont {Lee}, \citenamefont {Wallenstein},\ and\
  \citenamefont {Wieman}}]{HaenschPRL1975}%
  \BibitemOpen
  \bibfield  {author} {\bibinfo {author} {\bibfnamefont {T.}~\bibnamefont
  {H\"{a}nsch}}, \bibinfo {author} {\bibfnamefont {S.}~\bibnamefont {Lee}},
  \bibinfo {author} {\bibfnamefont {R.}~\bibnamefont {Wallenstein}}, \ and\
  \bibinfo {author} {\bibfnamefont {C.}~\bibnamefont {Wieman}},\ }\href
  {\doibase 10.1103/PhysRevLett.34.307} {\bibfield  {journal} {\bibinfo
  {journal} {Physical Review Letters}\ }\textbf {\bibinfo {volume} {34}},\
  \bibinfo {pages} {307} (\bibinfo {year} {1975})}\BibitemShut {NoStop}%
\bibitem [{\citenamefont {Dickenson}\ \emph {et~al.}(2013)\citenamefont
  {Dickenson}, \citenamefont {Niu}, \citenamefont {Salumbides}, \citenamefont
  {Komasa}, \citenamefont {Eikema}, \citenamefont {Pachucki},\ and\
  \citenamefont {Ubachs}}]{Dickenson2013}%
  \BibitemOpen
  \bibfield  {author} {\bibinfo {author} {\bibfnamefont {G.~D.}\ \bibnamefont
  {Dickenson}}, \bibinfo {author} {\bibfnamefont {M.~L.}\ \bibnamefont {Niu}},
  \bibinfo {author} {\bibfnamefont {E.~J.}\ \bibnamefont {Salumbides}},
  \bibinfo {author} {\bibfnamefont {J.}~\bibnamefont {Komasa}}, \bibinfo
  {author} {\bibfnamefont {K.~S.~E.}\ \bibnamefont {Eikema}}, \bibinfo {author}
  {\bibfnamefont {K.}~\bibnamefont {Pachucki}}, \ and\ \bibinfo {author}
  {\bibfnamefont {W.}~\bibnamefont {Ubachs}},\ }\href {\doibase
  10.1103/PhysRevLett.110.193601} {\bibfield  {journal} {\bibinfo  {journal}
  {Physical Review Letters}\ }\textbf {\bibinfo {volume} {110}},\ \bibinfo
  {pages} {193601} (\bibinfo {year} {2013})}\BibitemShut {NoStop}%
\bibitem [{\citenamefont {Meyer}\ \emph {et~al.}(2000)\citenamefont {Meyer},
  \citenamefont {Bagayev}, \citenamefont {Baird}, \citenamefont {Bakule},
  \citenamefont {Boshier}, \citenamefont {Breitr\"{u}ck}, \citenamefont
  {Cornish}, \citenamefont {Dychkov}, \citenamefont {Eaton}, \citenamefont
  {Grossmann}, \citenamefont {H\"{u}bl}, \citenamefont {Hughes}, \citenamefont
  {Jungmann}, \citenamefont {Lane}, \citenamefont {Liu}, \citenamefont {Lucas},
  \citenamefont {Matyugin}, \citenamefont {Merkel}, \citenamefont {zu~Putlitz},
  \citenamefont {Reinhard}, \citenamefont {Sandars}, \citenamefont {Santra},
  \citenamefont {Schmidt}, \citenamefont {Scott}, \citenamefont {Toner},
  \citenamefont {Towrie}, \citenamefont {Tr\"{a}ger}, \citenamefont
  {Willmann},\ and\ \citenamefont {Yakhontov}}]{Meyer2000}%
  \BibitemOpen
  \bibfield  {author} {\bibinfo {author} {\bibfnamefont {V.}~\bibnamefont
  {Meyer}}, \bibinfo {author} {\bibfnamefont {S.}~\bibnamefont {Bagayev}},
  \bibinfo {author} {\bibfnamefont {P.}~\bibnamefont {Baird}}, \bibinfo
  {author} {\bibfnamefont {P.}~\bibnamefont {Bakule}}, \bibinfo {author}
  {\bibfnamefont {M.}~\bibnamefont {Boshier}}, \bibinfo {author} {\bibfnamefont
  {A.}~\bibnamefont {Breitr\"{u}ck}}, \bibinfo {author} {\bibfnamefont
  {S.}~\bibnamefont {Cornish}}, \bibinfo {author} {\bibfnamefont
  {S.}~\bibnamefont {Dychkov}}, \bibinfo {author} {\bibfnamefont
  {G.}~\bibnamefont {Eaton}}, \bibinfo {author} {\bibfnamefont
  {A.}~\bibnamefont {Grossmann}}, \bibinfo {author} {\bibfnamefont
  {D.}~\bibnamefont {H\"{u}bl}}, \bibinfo {author} {\bibfnamefont
  {V.}~\bibnamefont {Hughes}}, \bibinfo {author} {\bibfnamefont
  {K.}~\bibnamefont {Jungmann}}, \bibinfo {author} {\bibfnamefont
  {I.}~\bibnamefont {Lane}}, \bibinfo {author} {\bibfnamefont {Y.-W.}\
  \bibnamefont {Liu}}, \bibinfo {author} {\bibfnamefont {D.}~\bibnamefont
  {Lucas}}, \bibinfo {author} {\bibfnamefont {Y.}~\bibnamefont {Matyugin}},
  \bibinfo {author} {\bibfnamefont {J.}~\bibnamefont {Merkel}}, \bibinfo
  {author} {\bibfnamefont {G.}~\bibnamefont {zu~Putlitz}}, \bibinfo {author}
  {\bibfnamefont {I.}~\bibnamefont {Reinhard}}, \bibinfo {author}
  {\bibfnamefont {P.}~\bibnamefont {Sandars}}, \bibinfo {author} {\bibfnamefont
  {R.}~\bibnamefont {Santra}}, \bibinfo {author} {\bibfnamefont
  {P.}~\bibnamefont {Schmidt}}, \bibinfo {author} {\bibfnamefont
  {C.}~\bibnamefont {Scott}}, \bibinfo {author} {\bibfnamefont
  {W.}~\bibnamefont {Toner}}, \bibinfo {author} {\bibfnamefont
  {M.}~\bibnamefont {Towrie}}, \bibinfo {author} {\bibfnamefont
  {K.}~\bibnamefont {Tr\"{a}ger}}, \bibinfo {author} {\bibfnamefont
  {L.}~\bibnamefont {Willmann}}, \ and\ \bibinfo {author} {\bibfnamefont
  {V.}~\bibnamefont {Yakhontov}},\ }\href {\doibase
  10.1103/PhysRevLett.84.1136} {\bibfield  {journal} {\bibinfo  {journal}
  {Physical Review Letters}\ }\textbf {\bibinfo {volume} {84}},\ \bibinfo
  {pages} {1136} (\bibinfo {year} {2000})}\BibitemShut {NoStop}%
\bibitem [{\citenamefont {Rosenband}\ \emph {et~al.}(2008)\citenamefont
  {Rosenband}, \citenamefont {Hume}, \citenamefont {Schmidt}, \citenamefont
  {Chou}, \citenamefont {Brusch}, \citenamefont {Lorini}, \citenamefont
  {Oskay}, \citenamefont {Drullinger}, \citenamefont {Fortier}, \citenamefont
  {Stalnaker}, \citenamefont {Diddams}, \citenamefont {Swann}, \citenamefont
  {Newbury}, \citenamefont {Itano}, \citenamefont {Wineland},\ and\
  \citenamefont {Bergquist}}]{Rosenband2008}%
  \BibitemOpen
  \bibfield  {author} {\bibinfo {author} {\bibfnamefont {T.}~\bibnamefont
  {Rosenband}}, \bibinfo {author} {\bibfnamefont {D.~B.}\ \bibnamefont {Hume}},
  \bibinfo {author} {\bibfnamefont {P.~O.}\ \bibnamefont {Schmidt}}, \bibinfo
  {author} {\bibfnamefont {C.~W.}\ \bibnamefont {Chou}}, \bibinfo {author}
  {\bibfnamefont {A.}~\bibnamefont {Brusch}}, \bibinfo {author} {\bibfnamefont
  {L.}~\bibnamefont {Lorini}}, \bibinfo {author} {\bibfnamefont {W.~H.}\
  \bibnamefont {Oskay}}, \bibinfo {author} {\bibfnamefont {R.~E.}\ \bibnamefont
  {Drullinger}}, \bibinfo {author} {\bibfnamefont {T.~M.}\ \bibnamefont
  {Fortier}}, \bibinfo {author} {\bibfnamefont {J.~E.}\ \bibnamefont
  {Stalnaker}}, \bibinfo {author} {\bibfnamefont {S.~A.}\ \bibnamefont
  {Diddams}}, \bibinfo {author} {\bibfnamefont {W.~C.}\ \bibnamefont {Swann}},
  \bibinfo {author} {\bibfnamefont {N.~R.}\ \bibnamefont {Newbury}}, \bibinfo
  {author} {\bibfnamefont {W.~M.}\ \bibnamefont {Itano}}, \bibinfo {author}
  {\bibfnamefont {D.~J.}\ \bibnamefont {Wineland}}, \ and\ \bibinfo {author}
  {\bibfnamefont {J.~C.}\ \bibnamefont {Bergquist}},\ }\href {\doibase
  10.1126/science.1154622} {\bibfield  {journal} {\bibinfo  {journal} {Science
  (New York, N.Y.)}\ }\textbf {\bibinfo {volume} {319}},\ \bibinfo {pages}
  {1808} (\bibinfo {year} {2008})}\BibitemShut {NoStop}%
\bibitem [{\citenamefont {Coddington}\ \emph {et~al.}(2009)\citenamefont
  {Coddington}, \citenamefont {Swann}, \citenamefont {Nenadovic},\ and\
  \citenamefont {Newbury}}]{Coddington2009}%
  \BibitemOpen
  \bibfield  {author} {\bibinfo {author} {\bibfnamefont {I.}~\bibnamefont
  {Coddington}}, \bibinfo {author} {\bibfnamefont {W.~C.}\ \bibnamefont
  {Swann}}, \bibinfo {author} {\bibfnamefont {L.}~\bibnamefont {Nenadovic}}, \
  and\ \bibinfo {author} {\bibfnamefont {N.~R.}\ \bibnamefont {Newbury}},\
  }\href {\doibase 10.1038/nphoton.2009.94} {\bibfield  {journal} {\bibinfo
  {journal} {Nature Photonics}\ }\textbf {\bibinfo {volume} {3}},\ \bibinfo
  {pages} {351} (\bibinfo {year} {2009})}\BibitemShut {NoStop}%
\bibitem [{\citenamefont {Marian}\ \emph {et~al.}(2004)\citenamefont {Marian},
  \citenamefont {Stowe}, \citenamefont {Lawall}, \citenamefont {Felinto},\ and\
  \citenamefont {Ye}}]{Marian2004}%
  \BibitemOpen
  \bibfield  {author} {\bibinfo {author} {\bibfnamefont {A.}~\bibnamefont
  {Marian}}, \bibinfo {author} {\bibfnamefont {M.~C.}\ \bibnamefont {Stowe}},
  \bibinfo {author} {\bibfnamefont {J.~R.}\ \bibnamefont {Lawall}}, \bibinfo
  {author} {\bibfnamefont {D.}~\bibnamefont {Felinto}}, \ and\ \bibinfo
  {author} {\bibfnamefont {J.}~\bibnamefont {Ye}},\ }\href {\doibase
  10.1126/science.1105660} {\bibfield  {journal} {\bibinfo  {journal} {Science
  (New York, N.Y.)}\ }\textbf {\bibinfo {volume} {306}},\ \bibinfo {pages}
  {2063} (\bibinfo {year} {2004})}\BibitemShut {NoStop}%
\bibitem [{\citenamefont {Gerginov}\ \emph {et~al.}(2005)\citenamefont
  {Gerginov}, \citenamefont {Tanner}, \citenamefont {Diddams}, \citenamefont
  {Bartels},\ and\ \citenamefont {Hollberg}}]{Gerginov2005}%
  \BibitemOpen
  \bibfield  {author} {\bibinfo {author} {\bibfnamefont {V.}~\bibnamefont
  {Gerginov}}, \bibinfo {author} {\bibfnamefont {C.~E.}\ \bibnamefont
  {Tanner}}, \bibinfo {author} {\bibfnamefont {S.~A.}\ \bibnamefont {Diddams}},
  \bibinfo {author} {\bibfnamefont {A.}~\bibnamefont {Bartels}}, \ and\
  \bibinfo {author} {\bibfnamefont {L.}~\bibnamefont {Hollberg}},\ }\href
  {\doibase 10.1364/OL.30.001734} {\bibfield  {journal} {\bibinfo  {journal}
  {Optics Letters}\ }\textbf {\bibinfo {volume} {30}},\ \bibinfo {pages} {1734}
  (\bibinfo {year} {2005})}\BibitemShut {NoStop}%
\bibitem [{\citenamefont {Wolf}\ \emph {et~al.}(2009)\citenamefont {Wolf},
  \citenamefont {van~den Berg}, \citenamefont {Ubachs},\ and\ \citenamefont
  {Eikema}}]{Wolf2009}%
  \BibitemOpen
  \bibfield  {author} {\bibinfo {author} {\bibfnamefont {A.~L.}\ \bibnamefont
  {Wolf}}, \bibinfo {author} {\bibfnamefont {S.~A.}\ \bibnamefont {van~den
  Berg}}, \bibinfo {author} {\bibfnamefont {W.}~\bibnamefont {Ubachs}}, \ and\
  \bibinfo {author} {\bibfnamefont {K.~S.~E.}\ \bibnamefont {Eikema}},\ }\href
  {\doibase 10.1103/PhysRevLett.102.223901} {\bibfield  {journal} {\bibinfo
  {journal} {Physical Review Letters}\ }\textbf {\bibinfo {volume} {102}},\
  \bibinfo {pages} {223901} (\bibinfo {year} {2009})}\BibitemShut {NoStop}%
\bibitem [{\citenamefont {Kandula}\ \emph {et~al.}(2010)\citenamefont
  {Kandula}, \citenamefont {Gohle}, \citenamefont {Pinkert}, \citenamefont
  {Ubachs},\ and\ \citenamefont {Eikema}}]{Kandula2010}%
  \BibitemOpen
  \bibfield  {author} {\bibinfo {author} {\bibfnamefont {D.~Z.}\ \bibnamefont
  {Kandula}}, \bibinfo {author} {\bibfnamefont {C.}~\bibnamefont {Gohle}},
  \bibinfo {author} {\bibfnamefont {T.~J.}\ \bibnamefont {Pinkert}}, \bibinfo
  {author} {\bibfnamefont {W.}~\bibnamefont {Ubachs}}, \ and\ \bibinfo {author}
  {\bibfnamefont {K.~S.~E.}\ \bibnamefont {Eikema}},\ }\href {\doibase
  10.1103/PhysRevLett.105.063001} {\bibfield  {journal} {\bibinfo  {journal}
  {Physical Review Letters}\ }\textbf {\bibinfo {volume} {105}},\ \bibinfo
  {pages} {063001} (\bibinfo {year} {2010})}\BibitemShut {NoStop}%
\bibitem [{\citenamefont {Witte}\ \emph {et~al.}(2005)\citenamefont {Witte},
  \citenamefont {Zinkstok}, \citenamefont {Ubachs}, \citenamefont
  {Hogervorst},\ and\ \citenamefont {Eikema}}]{Witte2005}%
  \BibitemOpen
  \bibfield  {author} {\bibinfo {author} {\bibfnamefont {S.}~\bibnamefont
  {Witte}}, \bibinfo {author} {\bibfnamefont {R.~T.}\ \bibnamefont {Zinkstok}},
  \bibinfo {author} {\bibfnamefont {W.}~\bibnamefont {Ubachs}}, \bibinfo
  {author} {\bibfnamefont {W.}~\bibnamefont {Hogervorst}}, \ and\ \bibinfo
  {author} {\bibfnamefont {K.~S.~E.}\ \bibnamefont {Eikema}},\ }\href {\doibase
  10.1126/science.1106612} {\bibfield  {journal} {\bibinfo  {journal} {Science
  (New York, N.Y.)}\ }\textbf {\bibinfo {volume} {307}},\ \bibinfo {pages}
  {400} (\bibinfo {year} {2005})}\BibitemShut {NoStop}%
\bibitem [{\citenamefont {Cing\"{o}z}\ \emph {et~al.}(2012)\citenamefont
  {Cing\"{o}z}, \citenamefont {Yost}, \citenamefont {Allison}, \citenamefont
  {Ruehl}, \citenamefont {Fermann}, \citenamefont {Hartl},\ and\ \citenamefont
  {Ye}}]{Cingoz2012}%
  \BibitemOpen
  \bibfield  {author} {\bibinfo {author} {\bibfnamefont {A.}~\bibnamefont
  {Cing\"{o}z}}, \bibinfo {author} {\bibfnamefont {D.~C.}\ \bibnamefont
  {Yost}}, \bibinfo {author} {\bibfnamefont {T.~K.}\ \bibnamefont {Allison}},
  \bibinfo {author} {\bibfnamefont {A.}~\bibnamefont {Ruehl}}, \bibinfo
  {author} {\bibfnamefont {M.~E.}\ \bibnamefont {Fermann}}, \bibinfo {author}
  {\bibfnamefont {I.}~\bibnamefont {Hartl}}, \ and\ \bibinfo {author}
  {\bibfnamefont {J.}~\bibnamefont {Ye}},\ }\href {\doibase
  10.1038/nature10711} {\bibfield  {journal} {\bibinfo  {journal} {Nature}\
  }\textbf {\bibinfo {volume} {482}},\ \bibinfo {pages} {68} (\bibinfo {year}
  {2012})}\BibitemShut {NoStop}%
\bibitem [{\citenamefont {Diddams}\ \emph {et~al.}(2007)\citenamefont
  {Diddams}, \citenamefont {Hollberg},\ and\ \citenamefont
  {Mbele}}]{Diddams2007}%
  \BibitemOpen
  \bibfield  {author} {\bibinfo {author} {\bibfnamefont {S.~a.}\ \bibnamefont
  {Diddams}}, \bibinfo {author} {\bibfnamefont {L.}~\bibnamefont {Hollberg}}, \
  and\ \bibinfo {author} {\bibfnamefont {V.}~\bibnamefont {Mbele}},\ }\href
  {\doibase 10.1038/nature05524} {\bibfield  {journal} {\bibinfo  {journal}
  {Nature}\ }\textbf {\bibinfo {volume} {445}},\ \bibinfo {pages} {627}
  (\bibinfo {year} {2007})}\BibitemShut {NoStop}%
\bibitem [{\citenamefont {Bjorkholm}\ and\ \citenamefont
  {Liao}(1976)}]{Bjorkholm1976}%
  \BibitemOpen
  \bibfield  {author} {\bibinfo {author} {\bibfnamefont {J.}~\bibnamefont
  {Bjorkholm}}\ and\ \bibinfo {author} {\bibfnamefont {P.}~\bibnamefont
  {Liao}},\ }\href {\doibase 10.1103/PhysRevA.14.751} {\bibfield  {journal}
  {\bibinfo  {journal} {Physical Review A}\ }\textbf {\bibinfo {volume} {14}},\
  \bibinfo {pages} {751} (\bibinfo {year} {1976})}\BibitemShut {NoStop}%
\bibitem [{\citenamefont {Stalnaker}\ \emph {et~al.}(2010)\citenamefont
  {Stalnaker}, \citenamefont {Mbele}, \citenamefont {Gerginov}, \citenamefont
  {Fortier}, \citenamefont {Diddams}, \citenamefont {Hollberg},\ and\
  \citenamefont {Tanner}}]{Stalnaker2010}%
  \BibitemOpen
  \bibfield  {author} {\bibinfo {author} {\bibfnamefont {J.~E.}\ \bibnamefont
  {Stalnaker}}, \bibinfo {author} {\bibfnamefont {V.}~\bibnamefont {Mbele}},
  \bibinfo {author} {\bibfnamefont {V.}~\bibnamefont {Gerginov}}, \bibinfo
  {author} {\bibfnamefont {T.~M.}\ \bibnamefont {Fortier}}, \bibinfo {author}
  {\bibfnamefont {S.~A.}\ \bibnamefont {Diddams}}, \bibinfo {author}
  {\bibfnamefont {L.}~\bibnamefont {Hollberg}}, \ and\ \bibinfo {author}
  {\bibfnamefont {C.~E.}\ \bibnamefont {Tanner}},\ }\href {\doibase
  10.1103/PhysRevA.81.043840} {\bibfield  {journal} {\bibinfo  {journal}
  {Physical Review A}\ }\textbf {\bibinfo {volume} {81}},\ \bibinfo {pages}
  {043840} (\bibinfo {year} {2010})}\BibitemShut {NoStop}%
\bibitem [{\citenamefont {Baklanov}\ and\ \citenamefont
  {Chebotayev}(1977)}]{Baklanov1977}%
  \BibitemOpen
  \bibfield  {author} {\bibinfo {author} {\bibfnamefont {Y.~V.}\ \bibnamefont
  {Baklanov}}\ and\ \bibinfo {author} {\bibfnamefont {V.~P.}\ \bibnamefont
  {Chebotayev}},\ }\href {\doibase 10.1007/BF00900075} {\bibfield  {journal}
  {\bibinfo  {journal} {Applied Physics}\ }\textbf {\bibinfo {volume} {12}},\
  \bibinfo {pages} {97} (\bibinfo {year} {1977})}\BibitemShut {NoStop}%
\bibitem [{\citenamefont {Fendel}\ \emph {et~al.}(2007)\citenamefont {Fendel},
  \citenamefont {Bergeson}, \citenamefont {Udem},\ and\ \citenamefont
  {H\"{a}nsch}}]{Fendel2007}%
  \BibitemOpen
  \bibfield  {author} {\bibinfo {author} {\bibfnamefont {P.}~\bibnamefont
  {Fendel}}, \bibinfo {author} {\bibfnamefont {S.~D.}\ \bibnamefont
  {Bergeson}}, \bibinfo {author} {\bibfnamefont {T.}~\bibnamefont {Udem}}, \
  and\ \bibinfo {author} {\bibfnamefont {T.~W.}\ \bibnamefont {H\"{a}nsch}},\
  }\href {\doibase 10.1364/OL.32.000701} {\bibfield  {journal} {\bibinfo
  {journal} {Optics Letters}\ }\textbf {\bibinfo {volume} {32}},\ \bibinfo
  {pages} {701} (\bibinfo {year} {2007})}\BibitemShut {NoStop}%
\bibitem [{\citenamefont {Ozawa}\ and\ \citenamefont
  {Kobayashi}(2012)}]{Ozawa2012}%
  \BibitemOpen
  \bibfield  {author} {\bibinfo {author} {\bibfnamefont {A.}~\bibnamefont
  {Ozawa}}\ and\ \bibinfo {author} {\bibfnamefont {Y.}~\bibnamefont
  {Kobayashi}},\ }\href {\doibase 10.1103/PhysRevA.86.022514} {\bibfield
  {journal} {\bibinfo  {journal} {Physical Review A}\ }\textbf {\bibinfo
  {volume} {86}},\ \bibinfo {pages} {022514} (\bibinfo {year}
  {2012})}\BibitemShut {NoStop}%
\bibitem [{\citenamefont {Barmes}\ \emph {et~al.}(2012)\citenamefont {Barmes},
  \citenamefont {Witte},\ and\ \citenamefont {Eikema}}]{Barmes2012}%
  \BibitemOpen
  \bibfield  {author} {\bibinfo {author} {\bibfnamefont {I.}~\bibnamefont
  {Barmes}}, \bibinfo {author} {\bibfnamefont {S.}~\bibnamefont {Witte}}, \
  and\ \bibinfo {author} {\bibfnamefont {K.~S.~E.}\ \bibnamefont {Eikema}},\
  }\href {\doibase 10.1038/nphoton.2012.299} {\bibfield  {journal} {\bibinfo
  {journal} {Nature Photonics}\ }\textbf {\bibinfo {volume} {7}},\ \bibinfo
  {pages} {38} (\bibinfo {year} {2012})}\BibitemShut {NoStop}%
\bibitem [{\citenamefont {Marian}\ \emph {et~al.}(2005)\citenamefont {Marian},
  \citenamefont {Stowe}, \citenamefont {Felinto},\ and\ \citenamefont
  {Ye}}]{Marian2005}%
  \BibitemOpen
  \bibfield  {author} {\bibinfo {author} {\bibfnamefont {A.}~\bibnamefont
  {Marian}}, \bibinfo {author} {\bibfnamefont {M.~C.}\ \bibnamefont {Stowe}},
  \bibinfo {author} {\bibfnamefont {D.}~\bibnamefont {Felinto}}, \ and\
  \bibinfo {author} {\bibfnamefont {J.}~\bibnamefont {Ye}},\ }\href {\doibase
  10.1103/PhysRevLett.95.023001} {\bibfield  {journal} {\bibinfo  {journal}
  {Physical Review Letters}\ }\textbf {\bibinfo {volume} {95}},\ \bibinfo
  {pages} {023001} (\bibinfo {year} {2005})}\BibitemShut {NoStop}%
\bibitem [{\citenamefont {Chui}\ \emph {et~al.}(2005)\citenamefont {Chui},
  \citenamefont {Ko}, \citenamefont {Liu}, \citenamefont {Shy}, \citenamefont
  {Peng},\ and\ \citenamefont {Ahn}}]{Chui2005}%
  \BibitemOpen
  \bibfield  {author} {\bibinfo {author} {\bibfnamefont {H.-C.}\ \bibnamefont
  {Chui}}, \bibinfo {author} {\bibfnamefont {M.-S.}\ \bibnamefont {Ko}},
  \bibinfo {author} {\bibfnamefont {Y.-W.}\ \bibnamefont {Liu}}, \bibinfo
  {author} {\bibfnamefont {J.-T.}\ \bibnamefont {Shy}}, \bibinfo {author}
  {\bibfnamefont {J.-L.}\ \bibnamefont {Peng}}, \ and\ \bibinfo {author}
  {\bibfnamefont {H.}~\bibnamefont {Ahn}},\ }\href {\doibase
  10.1364/OL.30.000842} {\bibfield  {journal} {\bibinfo  {journal} {Optics
  Letters}\ }\textbf {\bibinfo {volume} {30}},\ \bibinfo {pages} {842}
  (\bibinfo {year} {2005})}\BibitemShut {NoStop}%
\bibitem [{\citenamefont {Meshulach}\ and\ \citenamefont
  {Silberberg}(1999)}]{Meshulach1999}%
  \BibitemOpen
  \bibfield  {author} {\bibinfo {author} {\bibfnamefont {D.}~\bibnamefont
  {Meshulach}}\ and\ \bibinfo {author} {\bibfnamefont {Y.}~\bibnamefont
  {Silberberg}},\ }\href {\doibase 10.1103/PhysRevA.60.1287} {\bibfield
  {journal} {\bibinfo  {journal} {Physical Review A}\ }\textbf {\bibinfo
  {volume} {60}},\ \bibinfo {pages} {1287} (\bibinfo {year}
  {1999})}\BibitemShut {NoStop}%
\bibitem [{\citenamefont {Meshulach}\ and\ \citenamefont
  {Silberberg}(1998)}]{Meshulach1998}%
  \BibitemOpen
  \bibfield  {author} {\bibinfo {author} {\bibfnamefont {D.}~\bibnamefont
  {Meshulach}}\ and\ \bibinfo {author} {\bibfnamefont {Y.}~\bibnamefont
  {Silberberg}},\ }\href {\doibase 10.1038/24329} {\bibfield  {journal}
  {\bibinfo  {journal} {Nature}\ }\textbf {\bibinfo {volume} {396}},\ \bibinfo
  {pages} {239} (\bibinfo {year} {1998})}\BibitemShut {NoStop}%
\bibitem [{\citenamefont {Stowe}\ \emph {et~al.}(2008)\citenamefont {Stowe},
  \citenamefont {Thorpe}, \citenamefont {Pe'er}, \citenamefont {Ye},
  \citenamefont {Stalnaker}, \citenamefont {Gerginov},\ and\ \citenamefont
  {Diddams}}]{Stowe2008a}%
  \BibitemOpen
  \bibfield  {author} {\bibinfo {author} {\bibfnamefont {M.}~\bibnamefont
  {Stowe}}, \bibinfo {author} {\bibfnamefont {M.}~\bibnamefont {Thorpe}},
  \bibinfo {author} {\bibfnamefont {A.}~\bibnamefont {Pe'er}}, \bibinfo
  {author} {\bibfnamefont {J.}~\bibnamefont {Ye}}, \bibinfo {author}
  {\bibfnamefont {J.}~\bibnamefont {Stalnaker}}, \bibinfo {author}
  {\bibfnamefont {V.}~\bibnamefont {Gerginov}}, \ and\ \bibinfo {author}
  {\bibfnamefont {S.}~\bibnamefont {Diddams}},\ }\href {\doibase
  10.1016/S1049-250X(07)55001-9} {\bibfield  {journal} {\bibinfo  {journal}
  {Advances In Atomic, Molecular, and Optical Physics}\ }\textbf {\bibinfo
  {volume} {55}},\ \bibinfo {pages} {1} (\bibinfo {year} {2008})}\BibitemShut
  {NoStop}%
\bibitem [{\citenamefont {Li}\ \emph {et~al.}(2009)\citenamefont {Li},
  \citenamefont {Zhou}, \citenamefont {Wang},\ and\ \citenamefont
  {Zhan}}]{Li2009}%
  \BibitemOpen
  \bibfield  {author} {\bibinfo {author} {\bibfnamefont {R.-B.}\ \bibnamefont
  {Li}}, \bibinfo {author} {\bibfnamefont {L.}~\bibnamefont {Zhou}}, \bibinfo
  {author} {\bibfnamefont {J.}~\bibnamefont {Wang}}, \ and\ \bibinfo {author}
  {\bibfnamefont {M.-S.}\ \bibnamefont {Zhan}},\ }\href {\doibase
  10.1016/j.optcom.2008.12.034} {\bibfield  {journal} {\bibinfo  {journal}
  {Optics Communications}\ }\textbf {\bibinfo {volume} {282}},\ \bibinfo
  {pages} {1340} (\bibinfo {year} {2009})}\BibitemShut {NoStop}%
\bibitem [{\citenamefont {Arimondo}\ \emph {et~al.}(1977)\citenamefont
  {Arimondo}, \citenamefont {Inguscio},\ and\ \citenamefont
  {Violino}}]{Arimondo1977}%
  \BibitemOpen
  \bibfield  {author} {\bibinfo {author} {\bibfnamefont {E.}~\bibnamefont
  {Arimondo}}, \bibinfo {author} {\bibfnamefont {M.}~\bibnamefont {Inguscio}},
  \ and\ \bibinfo {author} {\bibfnamefont {P.}~\bibnamefont {Violino}},\ }\href
  {\doibase 10.1103/RevModPhys.49.31} {\bibfield  {journal} {\bibinfo
  {journal} {Reviews of Modern Physics}\ }\textbf {\bibinfo {volume} {49}},\
  \bibinfo {pages} {31} (\bibinfo {year} {1977})}\BibitemShut {NoStop}%
\end{thebibliography}%

\end{document}